\documentclass[12pt,preprint]{aastex}
\usepackage[dvips]{epsfig,color}

\newcommand{\eg}{{\it e.g.,}}
\newcommand{\ie}{{\it i.e.,}}
\newcommand{\etal}{{\it et al.}}

\newcommand{\vrms}{\ensuremath{v_{\rm rms}}}
\newcommand{\psd}{\ensuremath{\rho/\vrms^{3}}}

\newcommand{\ignore}[1]{\relax}

\shorttitle{Nonextensive Velocity Distributions}
\shortauthors{Barnes \etal}

\slugcomment{Accepted for publication in ApJ}

\begin{document}

\title{Velocity Distributions from Nonextensive Thermodynamics}
\author{Eric I. Barnes}
\affil{Department of Physics, University of Wisconsin --- La Crosse,
La Crosse, WI 54601}
\email{barnes.eric@uwlax.edu}
\author{Liliya L. R. Williams}
\affil{Department of Astronomy, University of Minnesota, Minneapolis,
MN 55455}
\email{llrw@astro.umn.edu}
\author{Arif Babul\altaffilmark{1}}
\affil{Department of Physics \& Astronomy, University of Victoria, BC,
Canada}
\email{babul@uvic.ca}
\author{Julianne J. Dalcanton\altaffilmark{2}}
\affil{Department of Astronomy, University of Washington, Box 351580,
Seattle, WA 98195}
\email{jd@astro.washington.edu}

\altaffiltext{1}{Leverhulme Visiting Professor, Universities of Oxford
and Durham}
\altaffiltext{2}{Alfred P. Sloan Foundation Fellow}

\begin{abstract}

There is no accepted mechanism that explains the equilibrium
structures that form in collisionless cosmological N-body simulations.
Recent work has identified nonextensive thermodynamics as an
innovative approach to the problem.  The distribution function that
results from adopting this framework has the same form as for
polytropes, but the polytropic index is now related to the degree of
nonextensiveness.  In particular, the nonextensive approach can mimic
the equilibrium structure of dark matter density profiles found in
simulations.  We extend the investigation of this approach to the
velocity structures expected from nonextensive thermodynamics.  We
find that the nonextensive and simulated N-body \vrms\ distributions
do not match one another.  The nonextensive \vrms\ profile is either
monotonically decreasing or displays little radial variation, each of
which disagrees with the \vrms\ distributions seen in simulations.  We
conclude that the currently discussed nonextensive models require
further modifications in order to corroborate dark matter halo
simulations.

\end{abstract}


\section{Introduction}\label{intro}

Simulations of structure formation have become well-refined and
sophisticated over the past few decades.  Typically, N-body
simulations follow collisionless particles that represent dark matter
and hierarchically form gravitationally bound structures.  These
structures have several ``universal'' characteristics, among them
their self-similar density profiles \citep[\eg][]{nfw96,m98,p03}.
However, as of yet there is no accepted physical mechanism for
explaining these shared properties, although there is no lack of
hypotheses \citep[\eg][]{l00,n01,b05,lu06}.  This work further
examines the recently advanced suggestion that nonextensive
thermodynamics can be used to determine the equilibria
\citep{h05,l05,hm06,klk06}.  For simplicity, we deal only with
spherically symmetric systems in this work.

\subsection{Nonextensive Thermodynamics}

The lack of short-range forces in self-gravitating, collisionless
systems throws the adoption of standard thermodynamics into question.
Instead, \citet{ts88} has proposed a generalized themodynamical
approach for such systems based on the nonextensiveness of entropy;
\ie\ the entropy of a composite system is not simply the sum of the
subsystems' entropies.  In particular, the entropy of a composite
system is given by,
\begin{equation}\label{ent}
S_{A+B}=S_A+S_B+(1-q)S_A S_B,
\end{equation}
where $S$ is the entropy, $A$ and $B$ refer to subsystems, and $q$ is
a parameter that describes the degree of nonextensiveness.  When $q=1$
the situation is extensive and corresponds to standard thermodynamics.
Nonextensive thermodynamics (sometimes referred to as $q$-statistics)
has been designed to describe systems with long dynamical memories, as
in self-gravitating collisionless objects.  Although other
applications have been discussed by \citet{ts99}, specific
applications of nonextensive thermodynamics to astrophysical
situations have been discussed by \citet{pp93,aly93,h05,l05}, and most
recently by \citet[][hereafter KLK]{klk06}.  One of the more
interesting results of the nonextensive approach is that the
associated distribution function has the same form as for a polytrope
\citep{pp93}.  In this case, the analogue to the polytropic index is
related to the $q$ value (see \S~\ref{constk}).  \citet{hm06} appeal
to nonextensive thermodynamics to interpret the correlation between
density profile slope and velocity anisotropy observed in N-body
simulations.  KLK have found that the density profiles corresponding
to nonextensive distribution functions can be found to match those of
virialized dark matter halos formed in cosmological simulations.
These studies have demonstrated that nonextensive thermodynamics is an
attractive option for theoretically grounding the behavior seen in
computer simulations.  After a brief synopsis of relevant halo
properties, we will discuss our findings that nonextensive
thermodynamics, as it currently stands, does not provide such a
basis.

\subsection{Overview of Simulated Dark Matter Halo Properties}

The virialized structures (halos) that are formed in cosmological
N-body simulations have several properties that appear to be
``universal''.  Density profiles have nonpower law shapes where a
scalelength divides a steeply declining outer region from a less
steeply declining inner cusp.  Most previous works investigating halo
formation focus on explaining this kind of density profile.  However,
the velocity structures of the halos are also important.  Like the
density profiles, N-body halo velocity dispersion distributions
$\vrms(r)$ have nonpower law shapes which peak near the density
scalelength.  The relevance of \vrms\ can be illustrated by the
quantity \psd\, which is reminiscent of and shares the dimensionality
of the phase space density.  As first noted in \citet{tn01}, this
quantity follows a power law distribution with radius $\psd \propto
r^{-\alpha}$, where $\alpha$ is a constant that depends on the
particular density profile.  This behavior appears to be linked to the
physics of collisionless collapse, given that the same power laws
result from independent semi-analytical halo formation models, as
shown by \citet{a05}.  The velocity structures of dark matter halos
therefore provide additional leverage for understanding the physical
mechanisms that drive equilibrium halo formation.

\section{Nonextensive Isotropic Equilibria}\label{nie}

KLK demonstrate that the equilibrium density profiles of sufficiently
nonextensive systems with velocity isotropy describe the results of
N-body simulations well (see Figure 2 of KLK).  Their work maintains a
constant degree of nonextensiveness $q$.  To maintain continuity with
KLK, we will adopt their variable $\kappa=1/(1-q)$ as the measure of
nonextensiveness.  We first investigate the predicted velocity
dispersion for these models, and then we proceed to allow $\kappa(r)$ to
vary.

\subsection{Constant $\kappa$}\label{constk}

Starting with the nonextensive definition of entropy \citep{ts88},
\begin{equation}\label{tsalent}
S_{\kappa}=k_B \kappa \left( \sum_i p_i^{1-1/\kappa} -1 \right),
\end{equation}
where $k_B$ is Boltzmann's constant and the sum is over all accessible
states (each with probability $p_i$), one can extremize $S_{\kappa}$
under the constraints of constant total mass and energy to find the
distribution function, as in \citet{pp93}.  \citet{spl98} find that
the same nonextensive velocity distribution functions can be found
using a variation on Maxwell's derivation of the standard distribution
functions, independent of any assumptions about energy.  As
\citet{spl98} make clear, the nonextensive approach demands velocity
isotropy.


Assuming a spherical system and adopting the
Poisson equation as the link between density and potential, the
nonextensive distribution function is obtained \citep{pp93,klk06},
\begin{equation}\label{distf}
f^{\pm}(E_r)=B^{\pm}\left[ 1+\frac{E_r}{\kappa
\sigma^2}\right]^{-\kappa},
\end{equation}
where $E_r \equiv v^2/2-\Psi$ is the relative specific energy, $v^2/2$
is the specific kinetic energy, $\Psi=-\Phi+\Phi_0$ ($\Phi$ is the
specific potential energy and $\Phi_0$ is a constant), and $\sigma$ is
an energy normalization constant.  The `$\pm$' denotes that $\kappa$
can be positive or negative and the normalization constant $B$
reflects that choice.  Again, note that this distribution function is
exactly that for a polytrope \citep{pp93}.  The nonextensive
thermodynamic framework provides a physical meaning to the polytropic
index; it is a measure of nonextensiveness.  With this distribution
function [which is isotropic since $f=f(E_r)$], it can be shown that
the density is,
\begin{equation}\label{dens}
\frac{\rho^{\pm}}{\rho_0}=\left[1-\frac{\Psi}{\kappa
\sigma^2}\right]^{3/2-\kappa}
\end{equation}
and the rms velocity distribution is given by,
\begin{equation}\label{vdisp}
\frac{\vrms^{\pm}}{v_0}=\left[1-\frac{\Psi}{\kappa
\sigma^2}\right]^{1/2},
\end{equation}
where $\vrms \equiv <\hspace{-0.1cm}v^2\hspace{-0.1cm}>^{1/2}$.  We
note that Equation~\ref{vdisp} is reminiscent of the expression for
the outer halo velocity dispersion given in Equation 3 of
\citet{hmg04}.  It is straightforward to show that these functions
satisfy the Jeans equation for mechanical equilibrium.
Equation~\ref{dens} can be solved for $\Psi$ and combined with the
Poisson equation,
\begin{equation}\label{poisson}
\nabla^2 \Psi=-4\pi G \rho
\end{equation}
to form a second-order differential equation for $\rho$ that we solve
numerically.

We turn our attention here to the velocity profiles given by
Equation~\ref{dens}.  We will specifically consider profiles with
$\kappa<0$, like those discussed in KLK, for reasons discussed at the
end of \S~\ref{varyk}.  For large values of $|\kappa|$ (\ie\ as a
system becomes increasingly extensive), \vrms\ approaches a constant
as the density profile becomes isothermal, as expected.  For more
modest values of $\kappa<0$, \vrms\ is always a continuously
decreasing function of radius.  However, this is not what is seen in
the results of cosmological N-body simulations, for which \vrms\ has a
well defined peak near the scalelength of the halo density profile
$r_0$.

\begin{figure}
\plotone{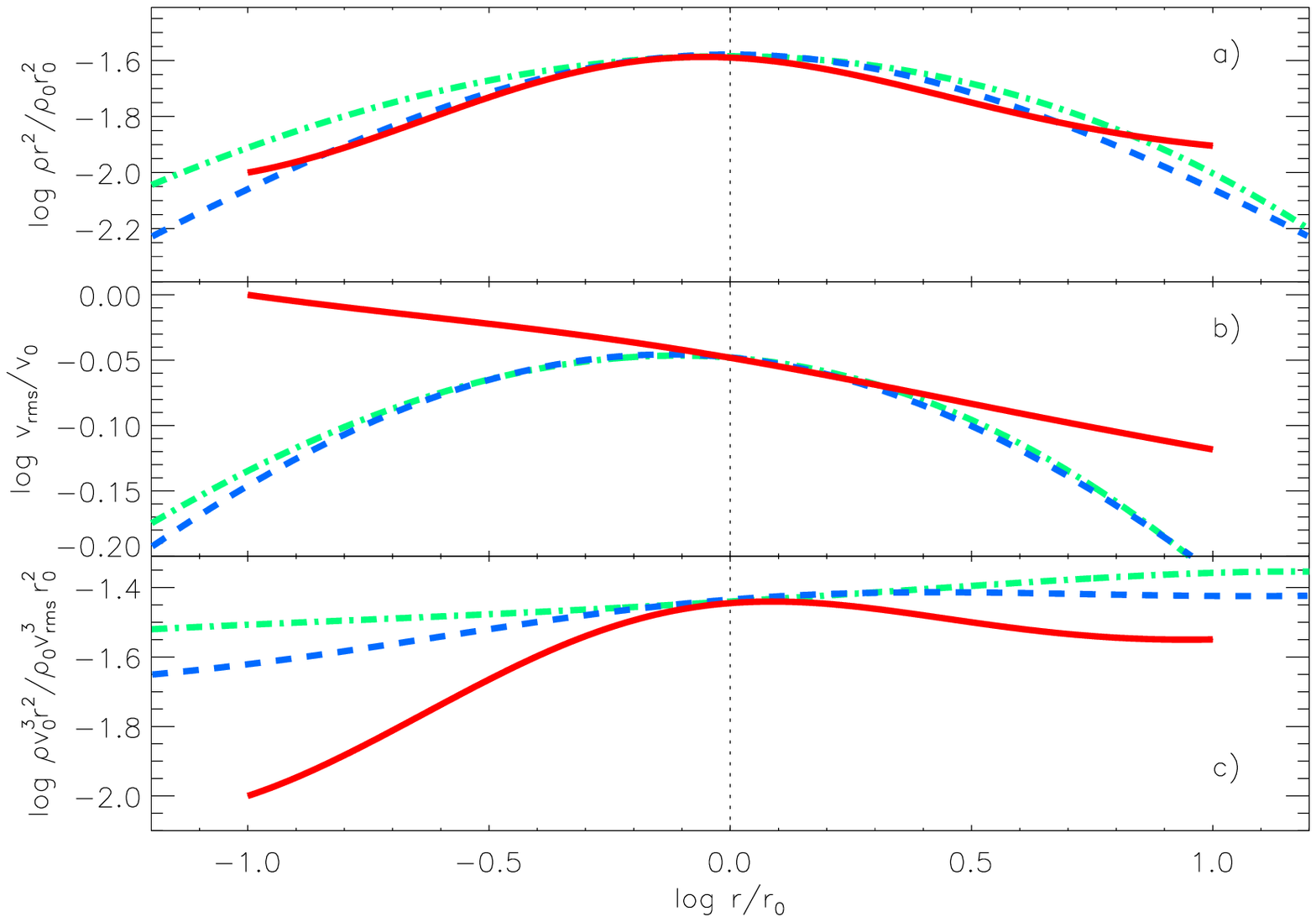}
\figcaption{The constant $\kappa$ nonextensive solutions for density
(panel a), velocity dispersion (panel b), and \psd\ (panel c) are
given by the solid red lines.  For comparison, corresponding curves
for the NFW (dashed blue lines) and N04 (dash-dotted green lines)
models are also shown.  The density and \psd\ profiles have been
scaled by $(r/r_0)^2$ to highlight variations from this power law
distribution.  The nonextensive \vrms\ distribution is monotonically
decreasing, in stark contrast to the behavior of the simulation-based
NFW and N04 models.  This discrepancy is also evident when comparing
the decidedly nonpower law shape of the solution \psd\ profile to the
very nearly power law empirical \psd\ profiles.  Note that the
discrepancy is evident at radii that are well-resolved by simulations.
\label{ckpsd}}
\end{figure}

Another view of this discrepancy is given by the phase-space
density-like quantity \psd.  The solid red lines in Figure~\ref{ckpsd}
represent nonextensive density (panel a), velocity dispersion (panel
b), and \psd\ profiles (panel c) using the KLK values $\kappa=-15$,
$\sigma=0.12$.  Note that the density and \psd\ distributions have
been scaled by $(r/r_0)^{-2}$ to highlight differences from this power
law profile.  Figure~\ref{ckpsd} also shows $\rho$, \vrms, and \psd\
profiles for isotropic \citet[][hereafter NFW]{nfw96} (dashed blue
lines) and \citet[][hereafter N04]{n04} (dash-dotted green lines)
models for comparison.  In the end, although a nonextensive, constant
$\kappa$ density profile can be found to mimic the density profiles
found in cosmological simulations, the corresponding velocity
structures do not match.

\subsection{Variable $\kappa$}\label{varyk}

We now relax the constraint of constant $\kappa$ and allow for
$\kappa(r)$ distributions.  As before, Jeans equilibrium is
maintained.  Varying $\kappa$ introduces derivatives of $\kappa$ to
Equation~\ref{poisson}, in addition to the derivatives of $\rho$,
allowing us to choose either: A) a $\rho$ distribution and solve for
$\kappa(r)$, or B) a $\kappa$ distribution and solve for $\rho(r)$.
We have tested our numerical implementations of these approaches for
self-consistency.

Accounting for radial variations of $\kappa$ is especially pertinent
given that it has recently been found that there is a direct
connection between $\kappa$ (or $q$) and the slope of a density
profile \citep{h05}.  The radial changes in density slope present in
dark matter halo models (\eg\ density slopes change from -1 to -3 as
radius increases in NFW profiles) then demand radial variation of
$\kappa$.

\begin{figure}
\plotone{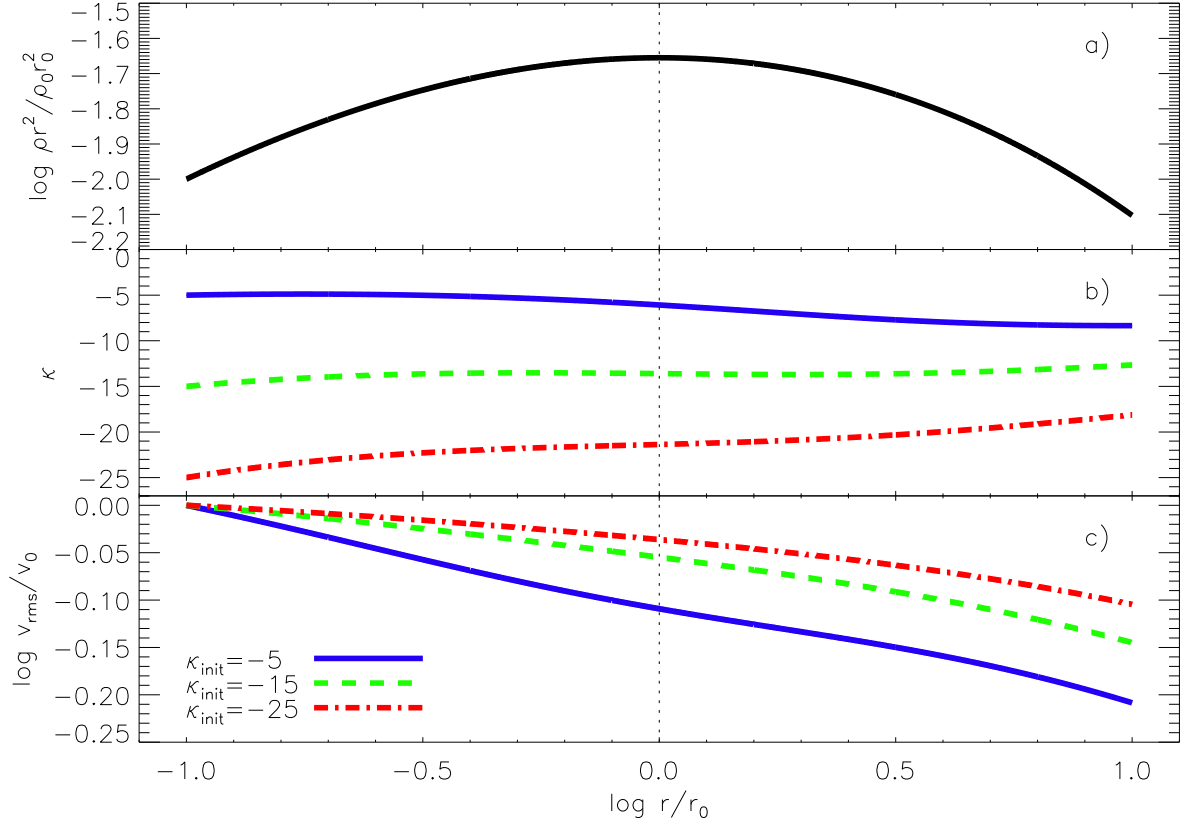}
\figcaption{When $\kappa$ is not constant, a $\kappa$ profile can be
found by solving Equation~\ref{poisson} with an assumed density
distribution.  The results shown here use the N04 density profile;
adopting an NFW profile leads to very similar results.  Panel a shows
the scaled N04 density distribution that is shared by all of the
solutions.  The solution $\kappa$ profiles (with $\kappa_{\rm
init}=0$) are shown in panel b.  Here, the solid blue, dashed green,
and dash-dotted red lines indicate solutions with $\kappa_{\rm
init}=-5,-15$, and $-25$, respectively.  The velocity dispersion
profiles (panel c) are montonically decreasing, as in the constant
$\kappa$ case.  Similarly, these \vrms\ profiles do not match those
from simulations.
\label{vksig}}
\end{figure}

Following track A dicussed above, both NFW and N04 density profiles
lead to derived $\kappa$ distributions that are always negative.
While we discuss the N04 profile in detail here, the NFW results are
essentially the same.  Figure~\ref{vksig} shows the scaled N04 density
profile (panel a), three $\kappa$ profiles (panel b), and the
resulting \vrms\ distributions (panel c).  The solid blue, dashed
green, and dash-dotted red lines represent solutions when $\kappa_{\rm
init}=-5,-15$, and $-20$, respectively.  We could not find solutions
with positive $\kappa$ values over any section of the distribution.
The resulting $\kappa$ distributions have roughly sinusoidal shapes
superimposed on linear trends which vary with the adopted initial
$\kappa$ value; for $\kappa_{\rm init} \ga -17$ the linear trends have
negative slopes, for $\kappa_{\rm init} \la -17$ the trends are
positive.  While these profiles are interesting in their own right,
the relevance to the current work is that the resulting velocity
dispersion profiles in Figure~\ref{vksig}c remain monotonically
decreasing functions, like the constant $\kappa$ case (solid line in
Figure~\ref{ckpsd}b).  We conclude that the nonextensive distribution
given in Equation~\ref{distf} does not reproduce all aspects of N-body
simulation results.

As track A fails to produce velocity dispersions similar to those in
simulations, we now discuss a $\kappa$ distribution designed to remedy
this problem.  We seek a function that is positive (negative) for
radii smaller (larger) than $r_0$.  This should give us a dispersion
profile that peaks near $r_0$.  Since $\kappa=0$ leads to infinite
entropy, we cannot simply transition from positive to negative values.
These arguments have led us to choose the following form,
\begin{equation}\label{cothk}
\kappa(r)=-A \coth[{s \log{(r/r_0)}}],
\end{equation}
where $\kappa \rightarrow A$ as $\log{r/r_0}\rightarrow -\infty$,
$\kappa \rightarrow -A$ as $\log{r/r_0}\rightarrow \infty$, and $s$
determines the rate of change between $A$ and the infinite value at
$\log{r/r_0}=0$.  As many functions can be created with similar
properties, we discuss this specific form only as an illustrative
example.

\begin{figure}
\plotone{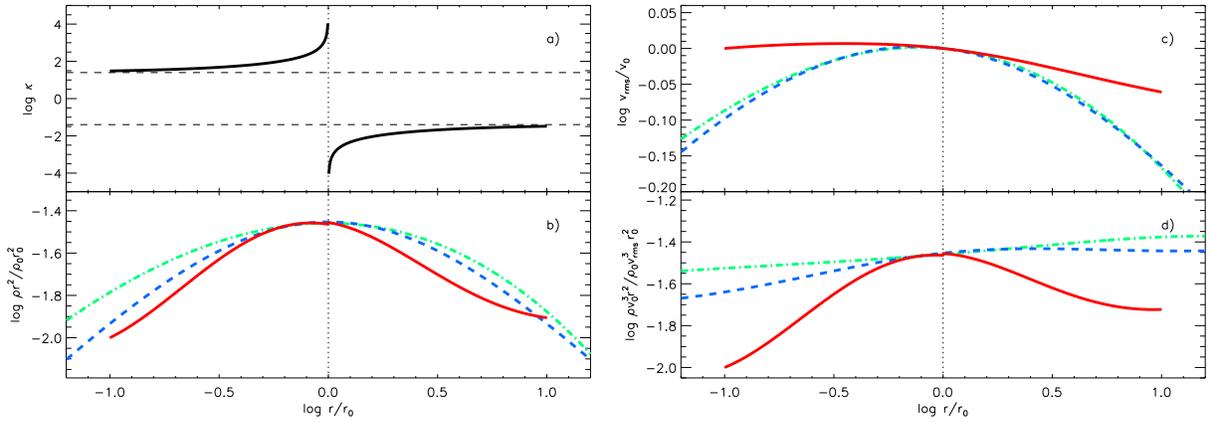}
\figcaption{For the $\kappa$ distribution described by
Equation~\ref{cothk} (panel a), we show the nonextensive solutions
(solid red lines) for density (panel b), velocity dispersion (panel
c), and \psd\ (panel d).  For comparison, corresponding curves for the
NFW (dashed blue lines) and N04 (dash-dotted green lines) models are
also shown.  As in Figure~\ref{ckpsd}, the densities and \psd\
profiles have been scaled by $(r/r_0)^2$ to highlight variations.  The
solution curve is a decent match to the empirical density profiles in
panel b.  Close inspection of panel c reveals that the solution \vrms\
profile does first increase before decreasing, in qualitative
agreement with the NFW and N04 curves.  However, it is quantitatively
very different from those curves, a fact also reflected in the
discrepancies apparent in the \psd\ profiles in panel d.
\label{vkpsd}}
\end{figure}

The $\kappa$ profile described by Equation~\ref{cothk} is shown in
Figure~\ref{vkpsd}a.  Where $\kappa<0$, we plot $-\log{|\kappa|}$.  A
density profile that reasonably approximates those from simulations
has the following parameter values; $A=25$, $s=0.5$, and $\sigma=0.12$
(Figure~\ref{vkpsd}b).  As in Figure~\ref{ckpsd}, the solid red,
dashed blue, and dash-dotted green lines represent solution, NFW, and
N04 profiles, respectively.  The \vrms\ distribution now has the
correct qualitative behavior; it is a rising function near the center
and decreases with increasing $r>r_0$ (Figure~\ref{vkpsd}c).  However,
it does not quantitatively match simulation results.  Specifically,
the \vrms\ profile is too flat.  This difference is also apparent when
looking at the \psd\ distributions in Figure~\ref{vkpsd}d.  The \psd\
profile from the nonextensive approach closely resembles the density,
which has a decidedly nonpower law shape.  There simply is not enough
variation in the \vrms\ distribution to compensate the changes in
$\rho$ and produce a power law.  Such discrepancies may reflect the
insufficiency of Equation~\ref{cothk}; further work may suggest more
appropriate $\kappa(r)$ expressions.

We note that this use of $\kappa >0$ for $\log{r/r_0}<0$ contradicts
the arguments made in KLK and \citet{l05}.  \citet{l05} points out
that negative $\kappa$ means: 1) that the entropy of a composite
system is less than in the extensive case, and 2) the heat capacity is
negative.  $\kappa>0$ implies a positive heat capacity, indicitive of
a system in which self-gravity is not important.  We do not speculate
further on the implications of this interpretation.  Positive $\kappa$
has been adopted simply to get the \vrms\ profile to have positive
slope at small $r$ like the simulation results.

\section{Summary \& Conclusions}\label{summary}

It has been recently shown that nonextensive thermodynamics can be
used to explain the density profiles of collisionless dark matter
halos formed in cosmological N-body simulations \citep[][KLK]{klk06}.
While this is an important step forward in understanding the physics
governing collisionless equilibria, N-body simulations also find
universal links between density and velocity distributions, such as
the power law behavior of \psd\ \citep[\eg][]{tn01}.  Can nonextensive
thermodynamics correctly predict the velocity behavior of systems
formed in cosmological simulations?

We find that a constant nonextensiveness parameter $\kappa$ (as in
KLK) predicts a velocity dispersion profile that is a continuously
decreasing function of radius, in conflict with simulated profiles.
We have extended the KLK approach to include radially varying $\kappa$
distributions designed to more closely match the velocity behavior of
simulation results.  While this approach produces qualitative matches
between the predictions and simulation results, there is no
quantitative agreement.  We describe four possible solutions to this
problem.

1) Equations~\ref{ent} and \ref{tsalent} correctly describe long-range
collisionless gravitational interactions, but cosmological N-body
simulations are not purely collisionless close to centers of halos due
to numerical effects. In this case, a velocity dispersion that
increases with radius (for $r<r_0$) may indicate that artificial
processes that mimic short-range interactions might be at play in the
simulations.  We do not speculate on what these might be \citep[but
see][]{ez05} or why they would lead to velocity dispersions increasing
with radius.  However, halos calculated from semi-analytical collapse
models can reproduce both the density and velocity profiles of N-body
halos, arguing against numerical effects.

2) The nonextensive approach of Equations~\ref{ent} and \ref{tsalent}
correctly predicts the global (asymptotic) equilibrium state,  but
N-body simulations (and possibly, real dark matter halos) settle to
nonlocal or quasi-equilibria through instabilities, such as the
radial orbit instability \citep{b05}.

3) Equations~\ref{ent} and \ref{tsalent} do not correctly describe
collisionless systems with long-range interactions, so a different
thermodynamic approach needs to be found to describe dark matter halo
simulations.

4) The approach taken in KLK and here does not account for velocity
anisotropy which is present in halos formed in cosmological N-body
simulations \citep{cl96,h99,b05,hm06,hs06}.  Preliminary work by the
authors that incorporates anisotropy into the nonextensive framework
suggests that this is not a panacea.  This adds to the evidence
presented in \citet{h06} that suggests the presence of anisotropy in
N-body halos requires that the distribution function have a more
complex form than that given by Equation~\ref{distf}, pointing
strongly towards the idea that a different thermodynamic approach must
be taken for these systems (point 3 above).

Regardless of the validity of these points, further investigation of
the relationship between nonextensive thermodynamics and
self-gravitating systems is vital to providing a physical
understanding, and independent corroboration, of the results of dark
matter halo formation simulations.

\acknowledgments
This work has been supported by NSF grant AST-0307604.  We thank
Thomas Kronberger, Manfred Leubner, and Eelco van Kampen for thoughful
comments.  We are also grateful to an anonymous referee who made
several helpful suggestions.  Research support for AB comes from the
Natural Sciences and Engineering Research Council (Canada) through the
Discovery grant program.  AB would also like to acknowledge support
from the Leverhulme Trust (UK) in the form of the Leverhulme Visiting
Professorship at the Universities of Oxford and Durham.  JJD was
partially supported through the Alfred P. Sloan Foundation.

\end{document}